\documentclass[12pt]{article}

\usepackage{hyperref}
\usepackage{authblk}
\usepackage{changepage}
\usepackage{gensymb}
\usepackage[utf8]{inputenc}
\usepackage{url}
\usepackage{graphicx}
\usepackage{tabulary}

\title{A reusable pipeline for large-scale fiber segmentation on unidirectional
fiber beds using fully convolutional neural networks}
\author[1,2]{Alexandre Fioravante de Siqueira\footnote{\href{mailto:alex.desiqueira@igdore.org}{alex.desiqueira@igdore.org}}}
\author[2]{Daniela M. Ushizima\footnote{\href{mailto:dushizima@lbl.gov}{dushizima@lbl.gov}}}
\author[1]{Stéfan J. van der Walt\footnote{\href{mailto:stefanv@berkeley.edu}{stefanv@berkeley.edu}}}

\affil[1]{\footnotesize \textit{Berkeley Institute for Data Science, University of California, Berkeley, USA}}

\affil[2]{\footnotesize \textit{Lawrence Berkeley National Laboratory, Berkeley, USA}}

\date{January, 2021}

\begin{document}

\maketitle

\begin{abstract}
Fiber-reinforced ceramic-matrix composites are advanced materials resistant to
high temperatures, with application to aerospace engineering. Their analysis
depends on the detection of embedded fibers, with semi-supervised techniques
usually employed to separate fibers within the fiber beds. Here we present an
open computational pipeline to detect fibers in ex-situ X-ray computed tomography fiber beds. To
separate the fibers in these samples, we tested four different architectures of
fully convolutional neural networks. When comparing our neural network approach to a semi-supervised one,
we obtained Dice and Matthews coefficients greater than $92.28 \pm 9.65\%$, reaching
up to $98.42 \pm 0.03 \%$, showing that the network results are close to the
human-supervised ones in these fiber beds, in some cases separating fibers that
human-curated algorithms could not find. The software we generated in this project
is open source, released under a permissive license, and can be freely adapted and
re-used in other domains. All data and instructions on how to download and use it are
also available.

\end{abstract}

\section{INTRODUCTION}

Fiber-reinforced ceramic-matrix composites are advanced materials used in
aerospace gas-turbine engines \cite{ZOK:2016, PADTURE:2016} and nuclear fusion
\cite{KOYANAGI:2018}, due to their resistance to temperatures 100–200 $\degree$C
higher than allows for the same applications.

Larson et al. investigated new manufacturing processes for curing preceramic
polymer into unidirectional fiber beds, studying the microstructure evolution
during matrix impregnation and aiming to reinforce ceramic-matrix composites
\cite{LARSON:2018, LARSON:2019}. They used X-ray computed tomography (CT) to
characterize the three-dimensional microstructure of their composites
non-destructively, studying their evolution \textit{in-situ} while processing
the materials at high temperatures \cite{LARSON:2018} and describing overall
fiber bed properties and microstructures of unidirectional composites
\cite{LARSON:2019}. The X-ray CT images acquired from these fiber beds are
available at Materials Data Facility \cite{BLAISZIK:2016}.

Larson et al.'s fiber beds have widths of approximately $1.5\,mm$, containing
5000–6200 fibers per stack. Each fiber has an average radius of $6.4 \pm 0.9\,\mu m$,
with diameters ranging from 13 to 20 pixels in the micrographs \cite{LARSON:2019}.
They present semi-supervised techniques to separate the fibers within the fiber
beds; their segmentation is available for five samples \cite{LARSON:2019_dataset}.
However, we considered their results could be improved using different techniques.
This motivated us to test alternative solutions.

In this study we separate fibers in \textit{ex-situ} X-ray CT fiber beds
of nine samples from Larson et al. The samples we used in this study
correspond to two general states: wet — obtained after pressure removal — and
cured. These samples were acquired using microtomographic instruments from the
Advanced Light Source at Lawrence Berkeley National Laboratory operated in a
low-flux, two-bunch mode \cite{LARSON:2019}. We used their reconstructions
obtained without phase retrieval; Larson et al. provide segmentations for five
of these samples \cite{LARSON:2019_dataset}, which we compare to our results.

To separate the fibers in these samples, we tested four different fully convolutional
neural networks (CNN, section \ref{methods:neural_nets}), algorithms from computer
vision and deep learning. When comparing our neural network approach to Larson
et al. results, we obtained Dice \cite{DICE:1945} and Matthews \cite{MATTHEWS:1975}
coefficients greater than $92.28 \pm 9.65\%$, reaching up to $98.42 \pm 0.03\%$,
showing that the network results are close to the human-supervised ones in these
fiber beds, in some cases separating fibers that the algorithms created by
\cite{LARSON:2019} could not find. All software and data generated in this study
are available for download. Instructions are given for downloading the data
and using the software. The code is open source, released under a
permissive software license, and can be adapted easily for other domains. 

\section{RESULTS}

\begin{figure}[hbt].
    \centering
    \includegraphics[width=0.8\linewidth]{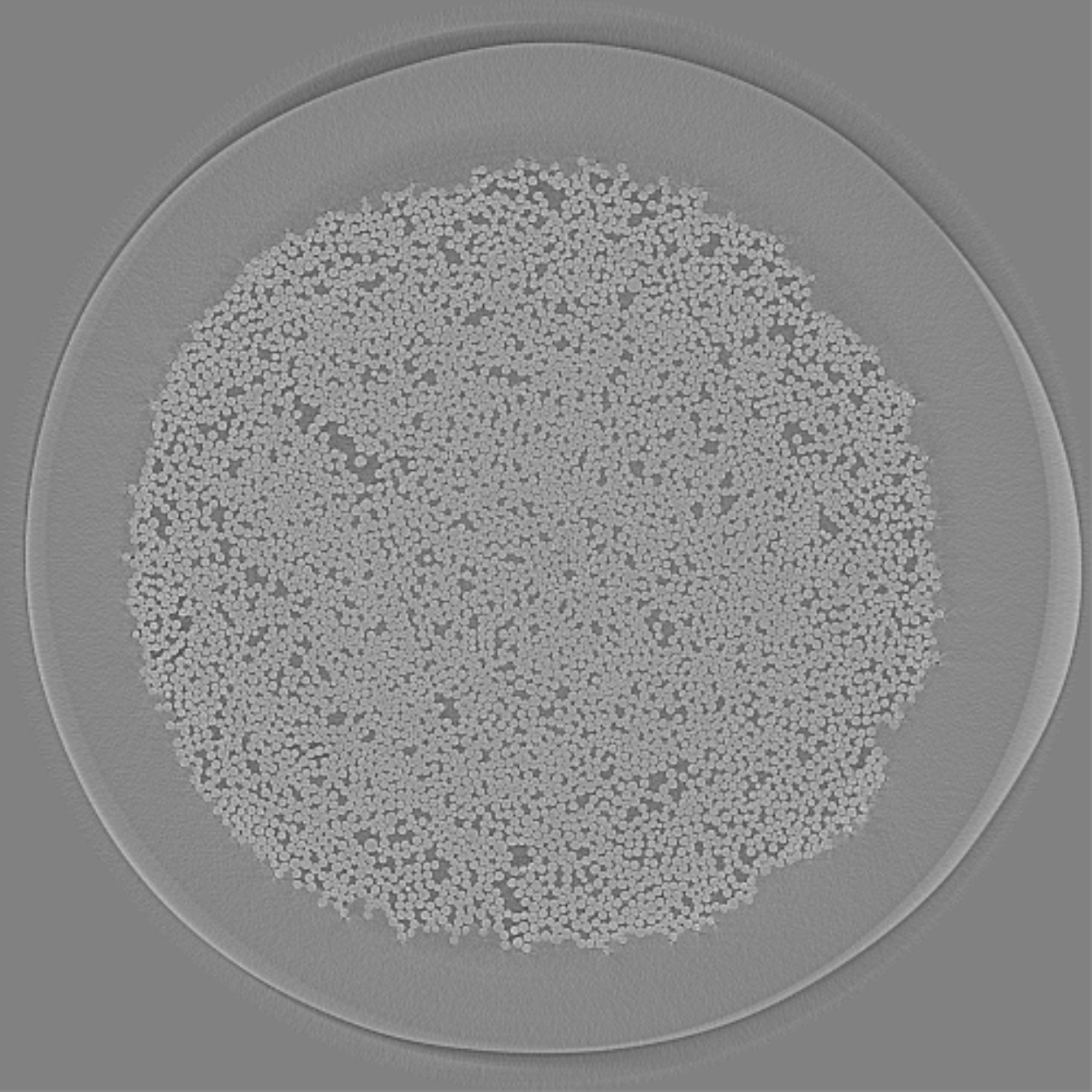}
    \caption{Slice number 1000 from the sample ``232p3 wet'', provided in
    \cite{LARSON:2019_dataset}. The whole sample contains 2160 slices. This
    slice represents the structure of the samples we processed: they contain
    the fiber bed (large circular structure) and the fibers within it (small
    round elements).}
    \label{fig:original_larson}
\end{figure}

Larson et al. provide segmentations for their fibers (Fig~\ref{fig:original_larson})
in five of the wet and cured samples, obtained using the following pipeline
\cite{LARSON:2019}:

\begin{enumerate}
    \item Fiber detection using the circular Hough transform \cite{YUEN:1989,
    ATHERTON:1999};
    \item Correction of improperly identified pixels using filters based on
    connected region size and pixel value, and by comparisons using ten slices
    above and below the slice of interest;
    \item Separation of fibers using the watershed algorithm \cite{MEYER:1994}.
\end{enumerate}

\begin{figure}[hbt].
    \centering
    \includegraphics[width=0.7\linewidth]{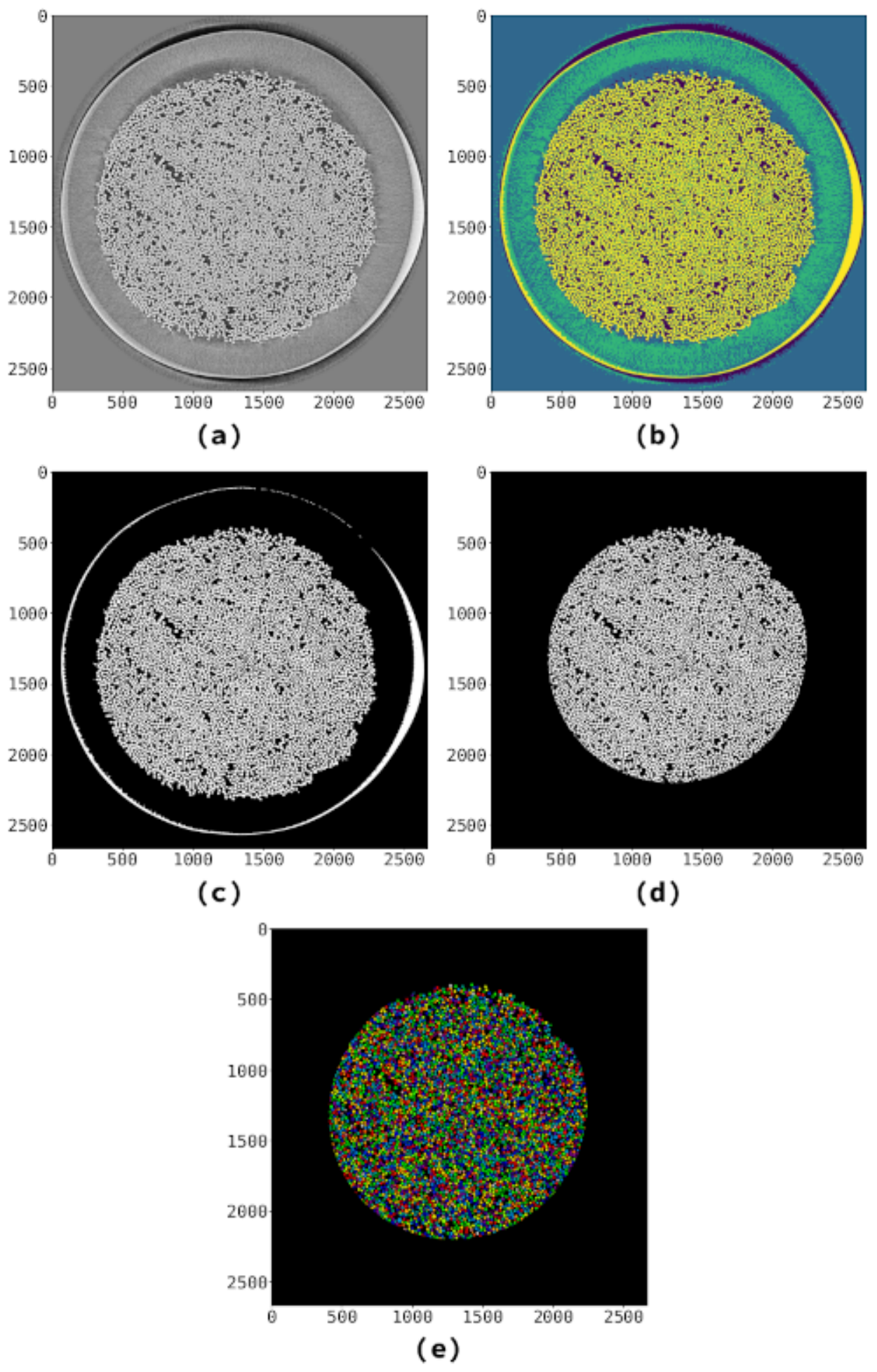}
    \caption{Rendering fibers detected in the limited region of interest by the
    classic pipeline. We exemplify the classic image processing pipeline using Fig
    \ref{fig:original_larson} as the input image. This solution presented
    limitations when processing fibers on the edfes of fiber beds. \textbf{(a)}
    Histogram equalization and TV Chambolle's filtering (parameter:
    \texttt{weight=0.3}). \textbf{(b)} Multi Otsu's resulting regions
    (parameter: \texttt{classes=4}). Fibers are located within the fourth region
    (in yellow). \textbf{(c)} Binary image obtained considering region four in
    (b) as the region of interest, and the remaining regions as the background.
    \textbf{(d)} the processed region from (c), as shown in Fig
    \ref{fig:original_larson}. \textbf{(e)} Regions resulting from the
    application of WUSEM on the region shown in (d) (parameters:
    \texttt{initial\_radius=0}, \texttt{delta\_radius=2},
    \texttt{watershed\_line=True}). Colormaps: (a, c, d) \texttt{gray}, (b)
    \texttt{viridis}, (e) \texttt{nipy\_spectral}.}
    \label{fig:classic_pipeline}
\end{figure}

However, their proposed method briefly describes these steps. There are no details
on parameters used, or the source code for their segmentation. We tried
different approaches to reproduce their results, focusing on separating the
fibers in the fiber bed samples. Our first approach was to create a classic,
unsupervised image processing pipeline. We used histogram equalization \cite{WOODS:1981},
Chambolle's total variation denoising \cite{RUDIN:1992, CHAMBOLLE:2004},
multi-Otsu threshold \cite{OTSU:1979, LIAO:2001}, and the WUSEM algorithm
\cite{DESIQUEIRA:2019} to separate each single fiber. The result is a labeled
image containing the separated fibers (Fig~\ref{fig:classic_pipeline}).
The pipeline presented limitations when processing fibers on the edges of fiber
beds, not being equivalent to the solution presented by Larson et al. We
restricted the segmentation region to have a satisfactory result
(Fig~\ref{fig:classic_pipeline}(d)), but the number of detected fibers is
reduced.

To obtain more accurate results, we evaluated four fully convolutional
neural network architectures: Tiramisu \cite{JEGOU:2017} and U-net
\cite{RONNEBERGER:2015}, as well as their three-dimensional counterparts, 3D
Tiramisu and 3D U-net \cite{CICEK:2016}. We also investigated whether
three-dimensional networks generate better segmentation results, leveraging
the structure of the material.

\subsection{Fully convolutional neural networks (CNN) for fiber detection}
\label{results:fully_convnets}

We implemented four architectures of fully convolutional neural networks (CNN) —
Tiramisu, U-net, 3D Tiramisu, and 3D U-net — to reproduce the results
provided by Larson et al. Labeled data, in our case,
consists of fibers within fiber beds. To train the neural networks to
recognize these fibers, we used slices from two different samples: 232p3 wet and
232p3 cured, registered according to the wet sample. Larson et al. provided the
fiber segmentation for these samples \cite{LARSON:2019_dataset}, which we used
as labels in the training. The training and validation datasets contained 250
and 50 images from each sample, respectively, in a total of 600 images. Each image
from the original samples have width and height size of $2560 \times 2560$ pixels.

\begin{figure}[hbt].
    \centering
    \includegraphics[width=0.75\linewidth]{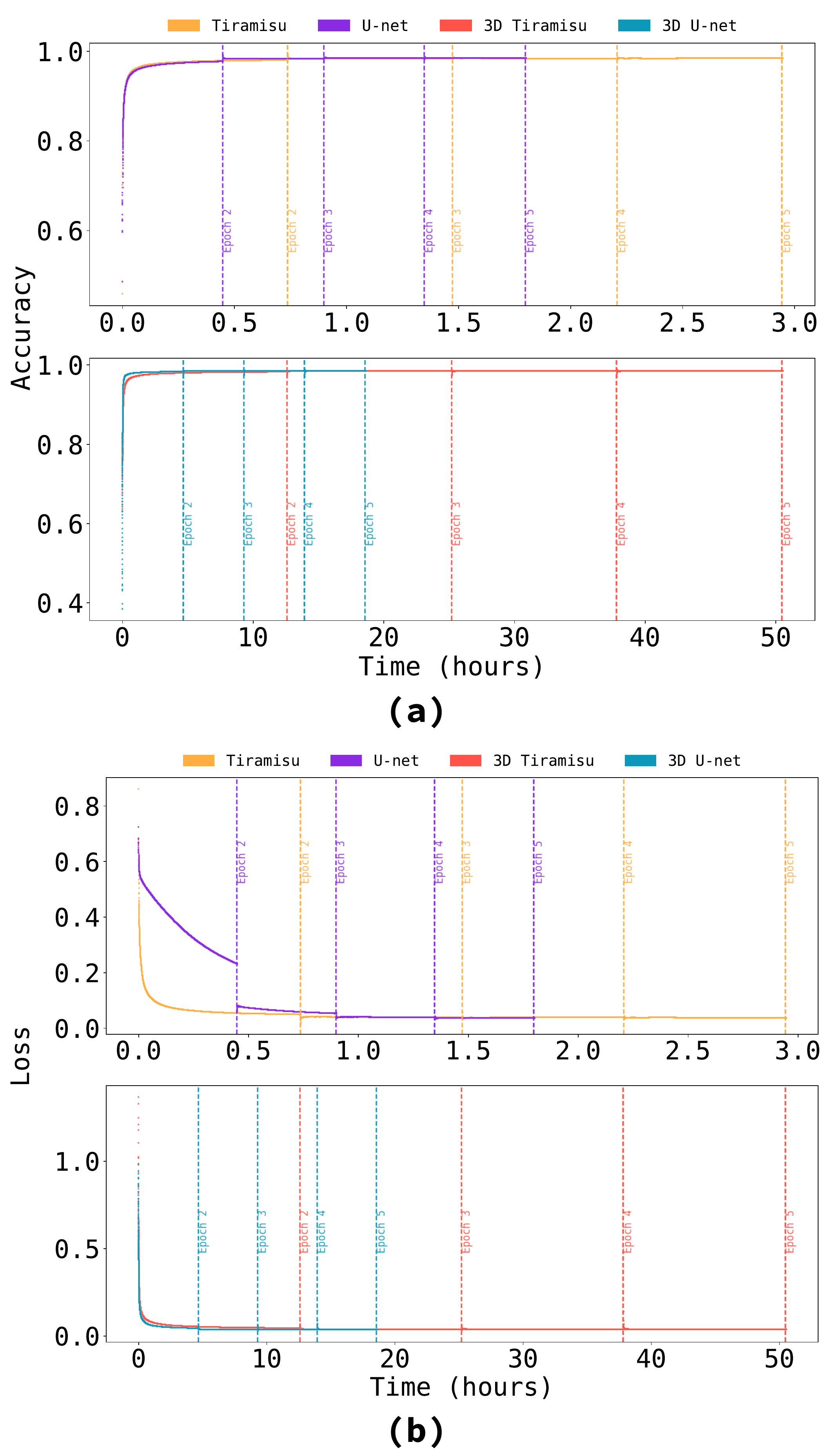}
    \caption{Accuracy \textbf{(a)} and loss \textbf{(b)} through time for each
    training epoch. All networks were trained during five epochs, reaching
    accuracy higher than 0.9 and loss lower than 0.1 on the first training
    epoch, except for the two-dimensional U-net. However, 2D U-net is the
    fastest to finish training, and reaches the lowest loss between the
    candidates. We attribute the subtle loss increase or accuracy decrease on
    the start of each epoch to the data augmentation process.}
    \label{fig:accuracy_loss}
\end{figure}

During the training procedure, the networks reached accuracy higher than 0.9 and
loss lower than 0.1 on the first epoch. Two-dimensional U-net is the exception,
presenting loss of 0.23 at the end of the first epoch. Despite that, 2D U-net
reaches the lowest loss between the four architectures at the end of its training.
2D U-net is also the fastest network to finish its training (7 h, 43 min),
followed by Tiramisu (13 h, 10 min), 3D U-net (24 h, 16 min) and 3D Tiramisu
(95 h, 49 min, Fig~\ref{fig:accuracy_loss}).

\begin{figure}[hbt].
    \centering
    \includegraphics[width=0.7\linewidth]{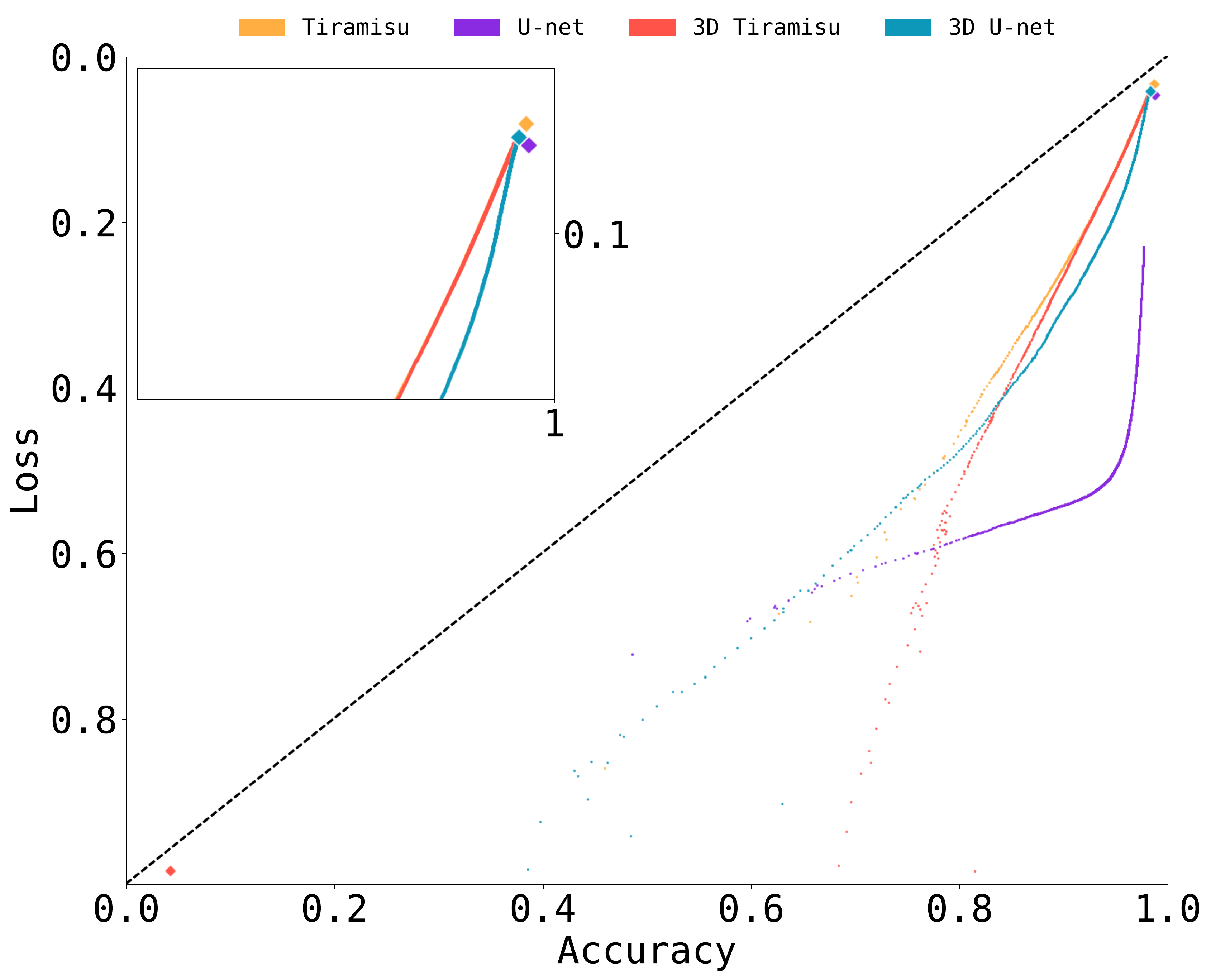}
    \caption{Accuracy vs. loss on the first epoch. Accuracy surpasses 0.9 and
    loss is lower than 0.1 for all networks during the first epoch, except for
    2D U-net (loss of 0.23). The large size of the training set and the
    similarities in the data are responsible for such numbers. Validation
    accuracy and validation loss on the first epoch are represented by diamonds.}
    \label{fig:accuracy_vs_loss}
\end{figure}

Still considering the 2D U-net, its convergence on the first epoch does not seem
as stable as the other networks (Fig~\ref{fig:accuracy_vs_loss}). However, this
does not impair U-net's accuracy (0.977 on the first epoch). Accuracy and loss
for the validation dataset also improve significantly from the first epoch:
Tiramisu had validation loss vs. validation accuracy ratio of 0.034 while U-net
had 0.048, 3D Tiramisu had 0.043, and 3D U-net had 0.043 as well. We attribute
these high accuracies and low losses to the large size of the training set and
the similarities between slices in the input data.

We used the coefficients from the training process to predict fibers in twelve
different datasets in total. These datasets were made available by Larson et al
\cite{LARSON:2019_dataset}, and we keep the same file identifiers for fast
cross-reference:

\begin{itemize}
    \item\textbf{``232p1'':} wet
    \item\textbf{``232p3'':} wet, cured, cured registered
    \item\textbf{``235p1'':} wet
    \item\textbf{``235p4'':} wet, cured, cured registered
    \item\textbf{``244p1'':} wet, cured, cured registered
    \item\textbf{``245p1'':} wet
\end{itemize}

Here, the first three numeric characters correspond to a material sample, and the
last character correspond to different extrinsic factors, e.g. deformation.
Despite being samples from similar materials, the reconstructed files presented
several differences, for example regarding amount of ringing artifacts, intensity
variation, noise, therefore they are considered as different samples in this paper.

\begin{figure}[hbt].
    \centering
    \includegraphics[width=0.9\linewidth]{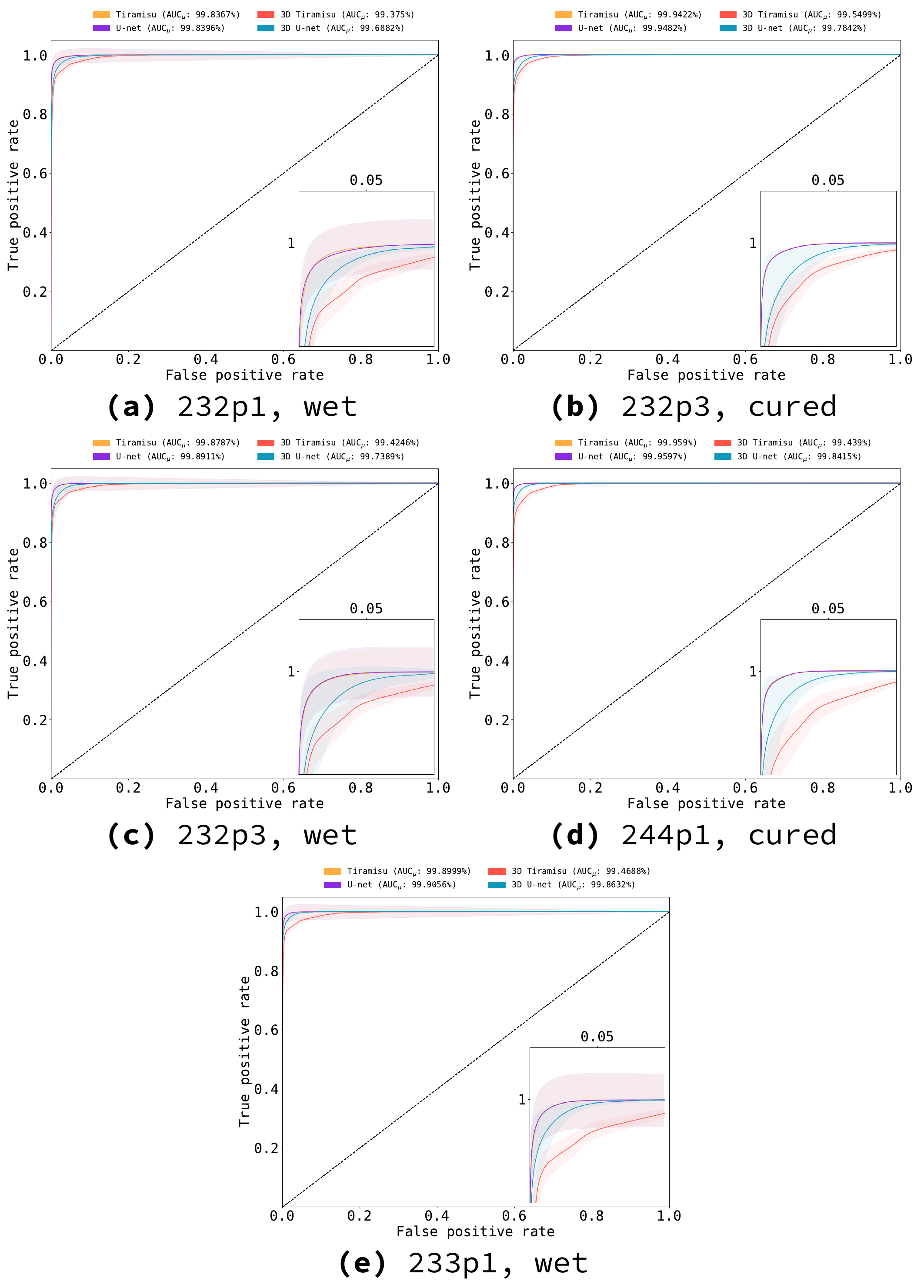}
    \caption{Mean and standard deviation for prediction times for each sample.
    As with processing, during training 2D U-net and 2D Tiramisu were the fastest
    architectures to process a sample in one hour, on average. 3D Tiramisu, being
    the slowest, takes in average more than a day to process one sample.}
    \label{fig:avg_predictions}
\end{figure}

We calculated the average processing time for each sample
(Fig~\ref{fig:avg_predictions}). The prediction time results are similar to the
training ones; 2D U-net and 2D Tiramisu are the fastest architectures to process
a sample, while 3D Tiramisu is the slowest.

\subsection{Evaluation of our results and comparison with Larson et al (2019)}
\label{results:eval_and_comp}

After processing all samples, we compared our predictions with the results that
Larson et al. made available on their dataset \cite{LARSON:2019_dataset}. They
provided five datasets from the twelve we processed: \textit{``232p1 wet''}, 
\textit{``232p3 cured''}, \textit{``232p3 wet''}, \textit{``244p1 cured''},
\textit{``244p1 wet''}.

\begin{figure}[hbt].
    \centering
    \includegraphics[width=0.9\linewidth]{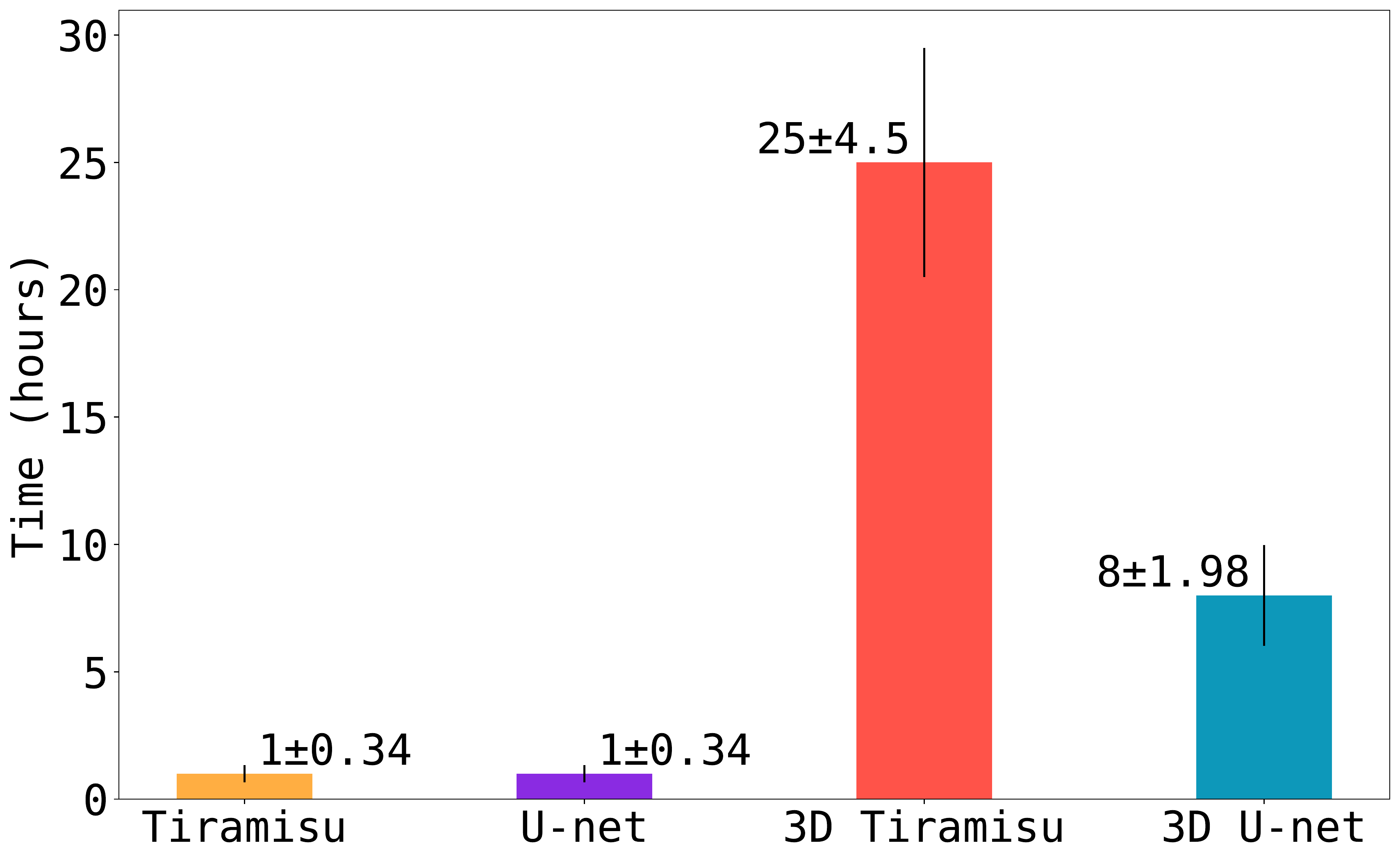}
    \caption{Receiver operating characteristic (ROC) and area under curve (AUC)
    from the comparison between the prediction for each network and the
    segmentation made available for five samples by Larson et al
    \cite{LARSON:2019_dataset}. ROC curves were calculated to all slices in a
    dataset; their mean areas and standard deviation intervals are presented.
    AUC is larger than 98\% in all comparisons, showing that our predictions
    are accurate when compared with Larson et al. semi-supervised method. The 2D
    versions of U-net and Tiramisu perform better when compared to their 3D
    alternatives.}
    \label{fig:roc_curves}
\end{figure}

First, we compared our predictions to their results using receiver operating
characteristic (ROC) curves and the area under curve (AUC,
Fig~\ref{fig:roc_curves}). AUC is larger than 98\% for all comparisons; therefore,
our predictions are accurate when compared with the semi-supervised method
suggested by Larson et al. The 2D versions of U-net and Tiramisu have similar
results, performing better than 3D U-net and 3D Tiramisu.

We also examined the binary versions of our predictions and compared them with
Larson et al. results. For each slice from the dataset, similarly to the volume,
we used a hard threshold of $0.5$; values above that are considered as fibers,
while values below that are treated as background. We used Dice \cite{DICE:1945}
and Matthews \cite{MATTHEWS:1975} correlation coefficients for our comparison
(\ref{tab:dice_and_matthews}). The comparison using U-net yields the highest
Dice and Matthews coefficients for three of five datasets. Tiramisu had highest
Dice/Matthews coefficients for the ``244p1, cured'' dataset, and both networks
have approximate results for ``232p1, wet''. 3D Tiramisu had the lowest
Dice and Matthews coefficients in our comparison.

\begin{table*}[hbt]  
    \centering
        \begin{adjustwidth}{-2.5cm}{}
        {\tiny
            \begin{tabulary}{1.5\textwidth}{LCCCCCCCC}
                \hline
                & \multicolumn{2}{c}{\textbf{Tiramisu}} & \multicolumn{2}{c}{\textbf{U-net}} & \multicolumn{2}{c}{\textbf{3D Tiramisu}} & \multicolumn{2}{c}{\textbf{3D U-net}} \\
                \hline
                \textbf{Sample} & \textbf{Dice} & \textbf{Matthews} & \textbf{Dice} & \textbf{Matthews} & \textbf{Dice} & \textbf{Matthews} & \textbf{Dice} & \textbf{Matthews} \\
                \hline
                \textbf{232p1, wet} & $97.58 \pm 2.29\%$ & $96.55 \pm 2.93\%$ & $97.58 \pm 2.20\%$ & $96.60 \pm 2.13\%$ & $94.54 \pm 6.73\%$ & $92.28 \pm 9.65\%$ & $95.59 \pm 0.74\%$ & $93.71 \pm 1.03\%$ \\
                \textbf{232p3, cured} & $98.21 \pm 0.04\%$ & $97.47 \pm 0.06\%$ & $98.26 \pm 0.04\%$ & $97.53 \pm 0.06\%$ & $95.25 \pm 6.36\%$ & $93.39 \pm 8.88\%$ & $95.90 \pm 1.00\%$ & $94.21 \pm 1.30\%$ \\
                \textbf{232p3, wet} & $97.79 \pm 2.15\%$ & $96.87 \pm 2.70\%$ & $97.85 \pm 2.12\%$ & $96.98 \pm 1.99\%$ & $94.86 \pm 6.90\%$ & $92.76 \pm 9.87\%$ & $95.68 \pm 1.97\%$ & $93.92 \pm 2.36\%$ \\
                \textbf{244p1, cured} & $98.42 \pm 0.03\%$ & $97.83 \pm 0.05\%$ & $98.38 \pm 0.04\%$ & $97.78 \pm 0.05\%$ & $94.55 \pm 7.74\%$ & $92.67 \pm 10.54\%$ & $96.30 \pm 1.25\%$ & $94.97 \pm 1.54\%$ \\
                \textbf{244p1, wet} & $98.08 \pm 2.53\%$ & $97.39 \pm 3.15\%$ & $98.10 \pm 2.39\%$ & $97.43 \pm 2.23\%$ & $94.81 \pm 7.81\%$ & $92.97 \pm 10.71\%$ & $96.67 \pm 1.00\%$ & $95.45 \pm 1.31\%$ \\
                \hline
            \end{tabulary}
        }
        \end{adjustwidth}
        \caption{Dice and Matthews coefficients for each sample, obtained from
        the comparison of our neural network results and data from Larson et al
        \cite{LARSON:2019_dataset}. U-net yields the highest Dice and Matthews
        coefficients for three of five samples. Tiramisu had highest
        Dice/Matthews coefficients for one of the datasets. 3D Tiramisu had the
        lowest Dice and Matthews coefficients.}
        \label{tab:dice_and_matthews}
    \end{table*}

\section{DISCUSSION}

The analysis of ceramic matrix composites (CMC) depends on the detection of its
fibers. Semi-supervised algorithms such as the one presented by Larson et al
\cite{LARSON:2019} can perform satisfactorily for that end. However, their
specific algorithm lack information on the parameters necessary for
replication. Reimplementing such methods without that information would lead to
inaccurate results, since the reported approach includes manual steps that
require human curation.

Convolutional neural networks are being used successfully in the segmentation
of different two- and three-dimensional scientific data (e.g., \cite{BANERJEE:2020,
TOKUOKA:2020, HORWATH:2020, MA:2020, SAITO:2019, LI:2018}), including microtomographies.
For example, fully convolutional
neural networks were used to generate 3D tau inclusion density maps \cite{ALEGRO:2020},
to segment the tidemark on osteochondral samples \cite{TIULPIN:2020}, and
3D models of structures of temporal-bone anatomy \cite{NIKAN:2020}.

Researchers are studying fiber-analysis detection for a while, using different
tools. There are several approaches using tracking, statistical approaches, or
classical image processing (e.g., \cite{CZABAJ:2014, BRICKER:2015, SENCU:2016,
USHIZIMA:2016, ZHOU:2016, EMERSON:2017, EMERSON:2018, CREVELING:2019}). To the
best of our knowledge, there are two different deep learning approaches for this problem:

\begin{itemize}
    \item Yu et al. \cite{YU:2018} use an unsupervised learning approach based
    on Faster R-CNN \cite{REN:2017} and a Kalman filter based tracking. They
    compare their results with Zhou et al. \cite{ZHOU:2016}, reaching a Dice
    coefficient of up to 99 \%.
    \item Miramontes et al. \cite{MIRAMONTES:2019} reach an average accuracy of
    93.75\% using a 2D LeNet-5 CNN \cite{LECUN:1998} to detect fibers in a
    specific sample.
\end{itemize}

Our study builds upon previous work by using similar material samples, but it expands
tests to many more samples as well as it includes the implemention and training of four architectures: 2D U-net,
2D Tiramisu, 3D U-net, and 3D Tiramisu, used to process twelve large datasets
($\approx$ 140 GB), and comparing our results with the gold standard data provided
by Larson et al. \cite{LARSON:2019_dataset} for five of them. We used ROC curves
and their area under curve (AUC) to ensure the quality of our predictions,
obtaining AUC larger than 98\% (Fig~\ref{fig:roc_curves}). Also, Dice and Matthews
coefficients were used to compare our results with Larson et al's solutions
(Table~\ref{tab:dice_and_matthews}), reaching coefficients of up to $98.42 \pm 0.03 \%$.

\begin{figure}[hbt].
    \centering
    \includegraphics[width=0.8\linewidth]{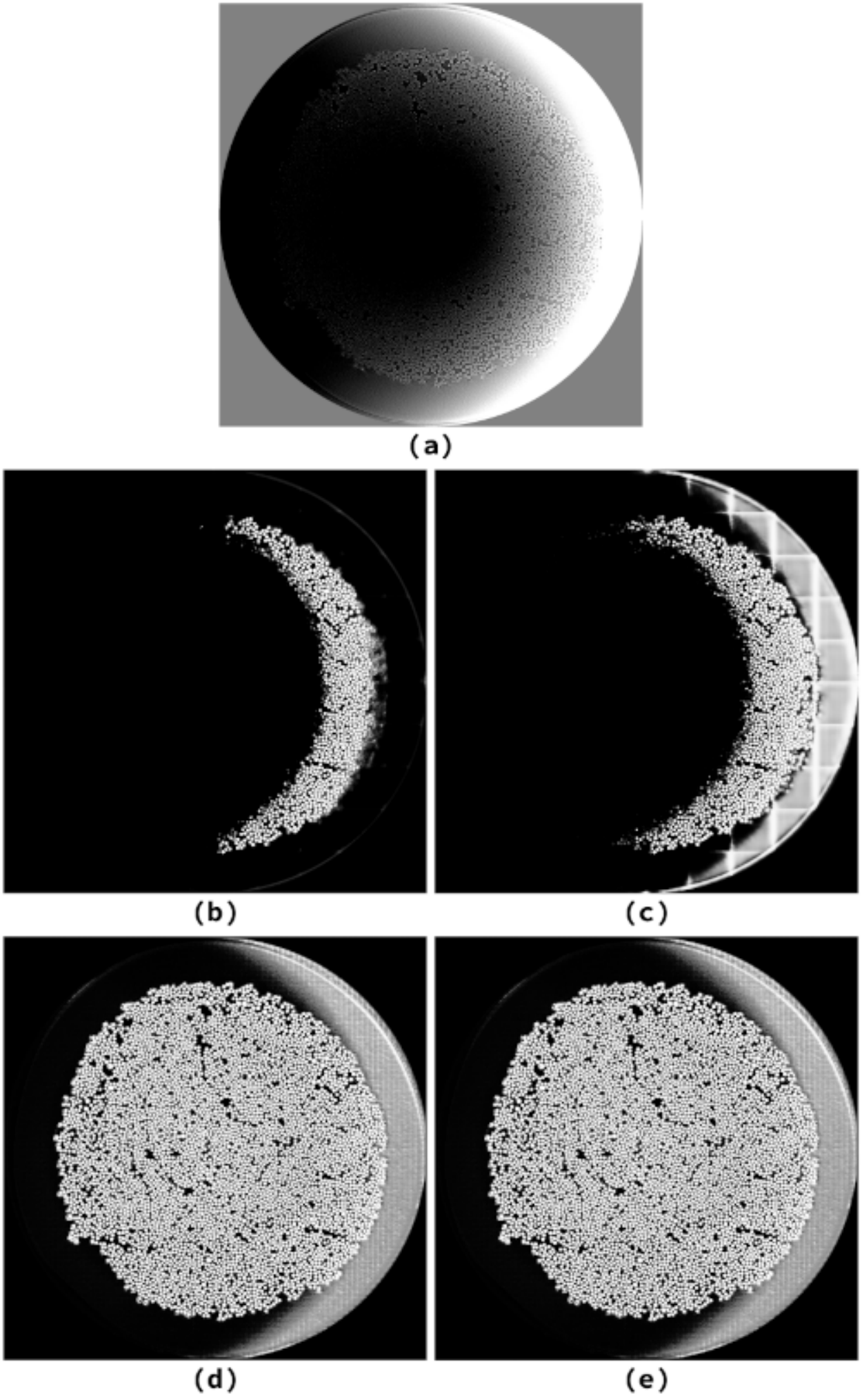}
    \caption{A defective slice on the sample ``232p3 wet'' and the segmentation
             resulting from each architecture. While the 2D architectures results
             are impaired by the defects present in the input image, the 3D ones
             leverage from the sample structure to present a better segmentation
             result. \textbf{(a)} Original defective image, \textbf{(b)} U-net
             prediction, \textbf{(c)} 3D U-net prediction, \textbf{(d)} Tiramisu
             prediction, \textbf{(e)} 3D Tiramisu prediction.}
    \label{fig:defect_performance}
\end{figure}

When processing a defective slice, the 3D architectures perform better when compared
to the 2D ones, since they leverage from information of the material structure
(Fig~\ref{fig:defect_performance}).

Based on our research, we recommend using the 2D U-net to process microtomographies
of CMC fibers. Both 2D networks lead to similar accuracy and loss values
(\ref{tab:dice_and_matthews}) in our comparisons; however, U-nets achieve these
numbers in a shorter time, when compared to Tiramisu.
The 3D architectures, while presenting the advantage of performing better in
defective samples (Fig~\ref{fig:defect_performance}), do not achieve general
results comparable to the 2D architectures. Satisfactory results using the 3D
architectures could be achieved with more training time; however, one should
notice that the training we proposed (Fig~\ref{fig:accuracy_loss}) and the
consequent predictions (Fig~\ref{fig:avg_predictions}) required a considerable
computing time, and more training could lead to marginal improvements when
compared to their 2D alternatives.

\begin{figure}[hbt].
    \centering
    \includegraphics[width=0.8\linewidth]{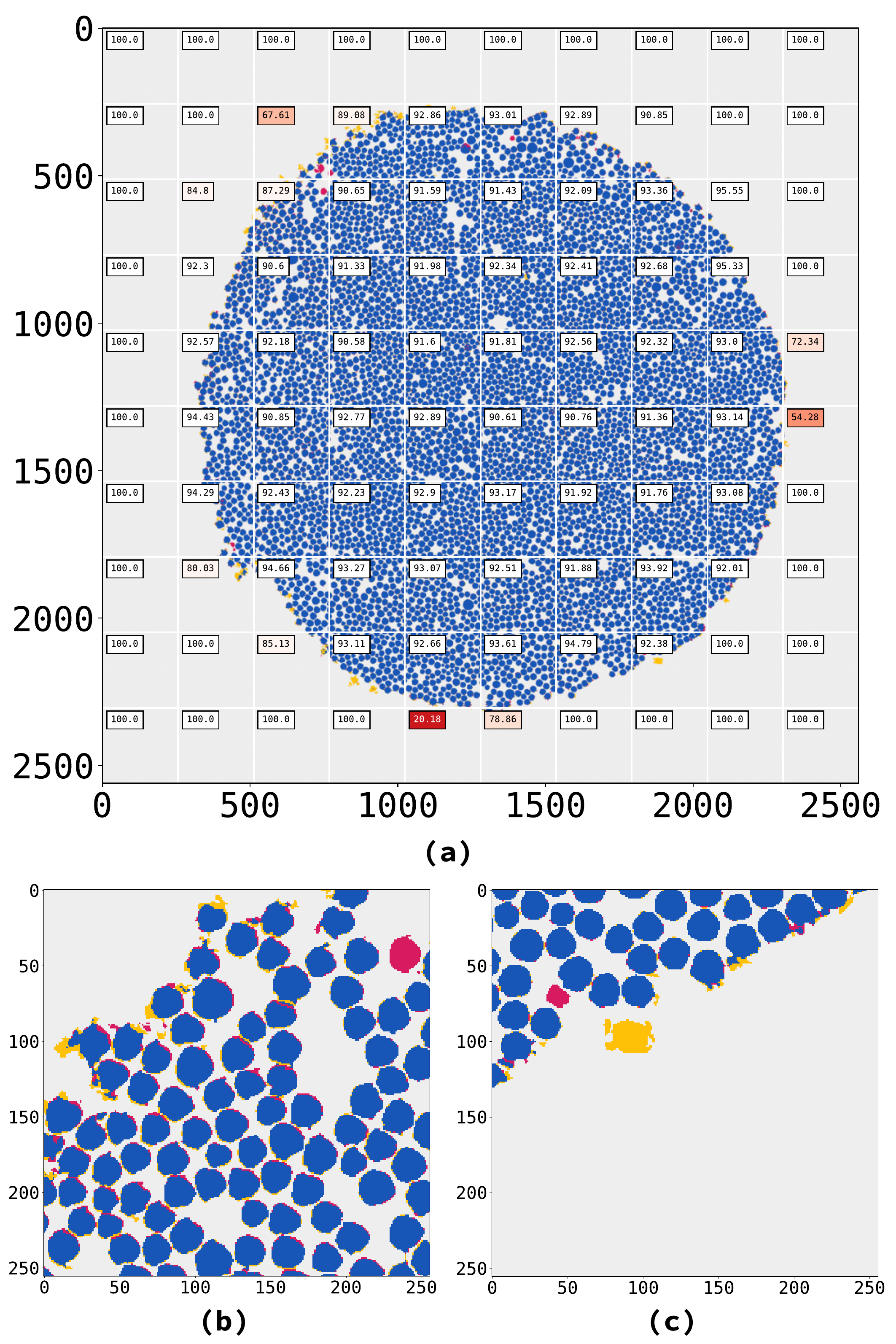}
    \caption{Visual comparison between 2D U-net and Larson et al. results
    for sample ``232p3 wet''. Each part of this image is obtained combining both
    ours and Larson et al.'s results; we compared each slice, and presented the
    ones that return the lowest Matthews comparison coefficient. Labels present
    the Matthews coefficient for each slice. \textbf{(b, c)} slices presenting
    fibers found only by U-net (in red), while some well-defined structures
    close to the borders are found only by Larson et al. (in yellow). Slice size:
    $256 \times 256$. Colors set according to the comparison. Blue: true positives;
    red: false positives; yellow: false negatives; gray: true negatives. For
    more details on the comparison coefficients, please see \ref{subsec:eval}.}
    \label{fig:visual_comp}
\end{figure}

Our CNN architectures perform to the level of human-curated accuracy — i.e.,
Larson et al. semi-supervised approach —, sometimes even surpassing Larson et
al. algorithm. For instance, the 2D U-net process fibers their algorithm
could not find (Fig~\ref{fig:visual_comp}).

\begin{figure}[hbt].
    \centering
    \includegraphics[width=1\linewidth]{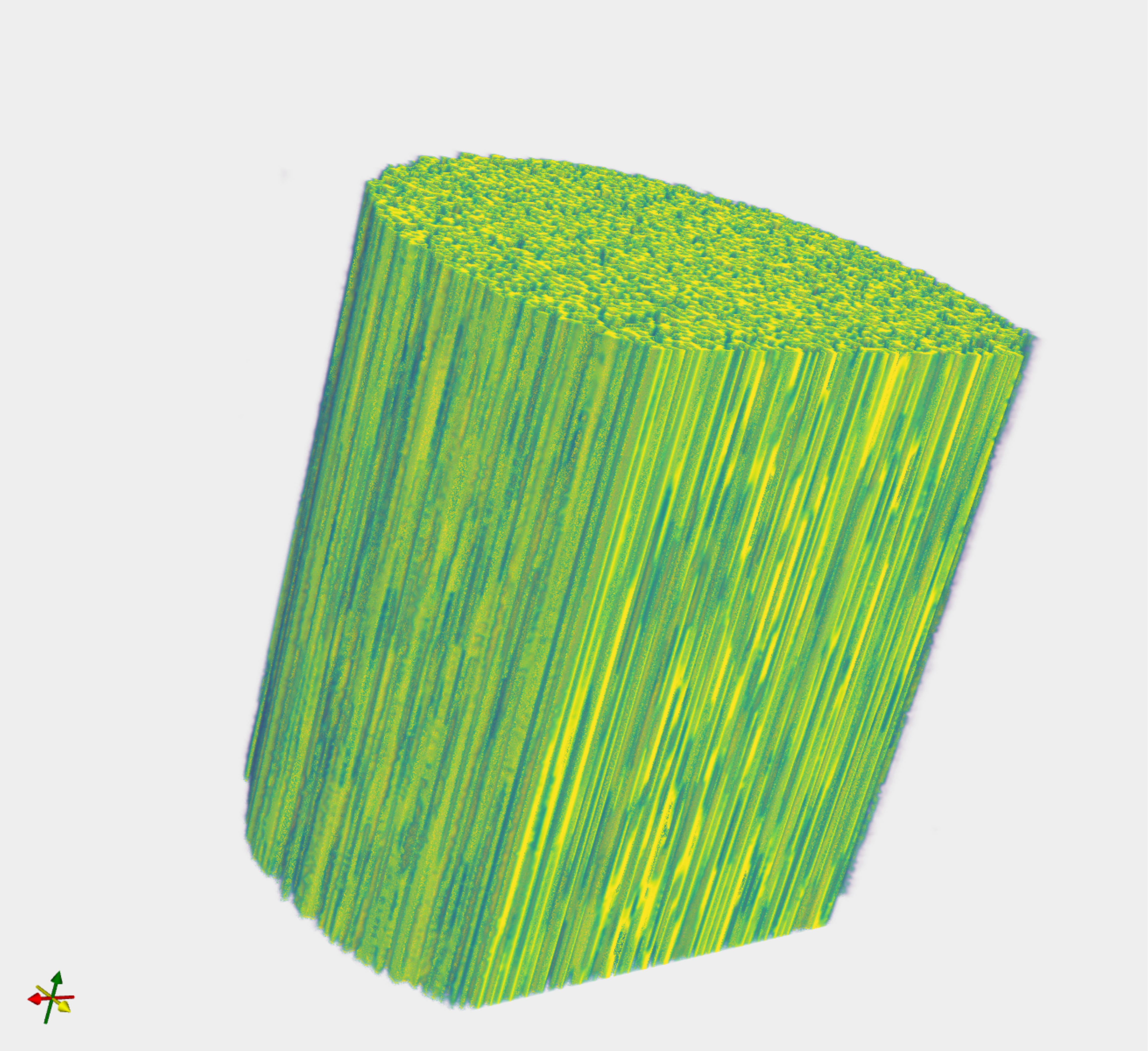}
    \caption{Fibers on the sample ``232p3 wet'' processed using the U-net
    architecture. As seen in the longitudinal cut, this pipeline identifies
    fibers across the sample height despite the absence of tracking.}
    \label{fig:3d_render}
\end{figure}

Using the data processed by the U-net architecture, we can render a
three-dimensional visualization of the fibers (Fig~\ref{fig:3d_render}). Despite
the absence of tracking, the segmentation provided by the U-net identifies the
fibers across the stack.

In this paper, we presented deep learning solutions to analyze microtomographies
of CMC fibers in fiber beds. The data used is publicly available \cite{LARSON:2019_dataset}
and was acquired in a real materials design experiment. Our solutions are
comparable to human-curated results, being capable of predicting fibers in large
stacks of microtomographies without human intervention.

Despite the encouraging results we achieved in this study, there is further
space for improvements. For example, we aim to study how an ensemble of our
trained networks would perform in these samples. Another proposal would be
to work with different thresholds at the last layer of your network. We maintained
a hard threshold of $0.5$, that suited the sigmoid on the last layer of the
CNN we implemented. We could also use conditional random field networks for
that end.

\section{METHODS}

\subsection{Fully convolutional neural networks}
\label{methods:neural_nets}

We implemented four architectures — two dimensional U-net \cite{RONNEBERGER:2015}
and Tiramisu \cite{JEGOU:2017}, and their three-dimensional versions — to attempt
reproducing the results provided by Larson et al. We used supervised algorithms:
they rely on labeled data to learn what are the regions of interest — in our case,
fibers within microtomographies of fiber beds.

All CNN algorithms were implemented using TensorFlow \cite{ABADI:2016} and Keras
\cite{CHOLLET:2015} on a computer with two Intel Xeon Gold processors
6134 and two Nvidia GeForce RTX 2080 graphical processing units. Each GPU has 10 GB of RAM.

To train the neural networks on how to recognize the fibers, we used slices
from two different samples: ``232p3 wet'' and ``232p3 cured'', registered according
to the wet sample. Larson et al. provided the fiber segmentation for these
samples, which we used as labels in the training. The training and validation
procedures processed 350 and 149 images from each sample, respectively; a total
of 998 images. Each image from the original samples have width and height size
of $2560 \times 2560$ pixels.

To feed the two-dimensional networks, we padded the images with 16 pixels, of value zero, in
each dimension. Then, each image was cut into tiles of size
$288 \times 288$, each 256 pixels, creating an overlap of 32 pixels. These
overlapping regions, which are again removed after processing, avoid
artifacts on the borders of processed tiles. Therefore, each input slice
generated 100 images with $288 \times 288$ pixels, in a total of 50,000 images
for the training set, and 10,000 for the validation set.

We needed to pre-process the training images differently to train the
three-dimensional networks. We loaded the entire samples, each with size
$2160 \times 2560 \times 2560$, and padded their dimensions with 16 pixels. Then,
we cut slices of size $64 \times 64 \times 64$ voxels, each 32 pixels. Hence,
the training and validation sets for the three-dimensional networks have 96,000
and 19,200 cubes, respectively.

We implemented data augmentation in our pipeline, aiming for a network capable of
processing samples with different characteristics. We augmented the images on the training
sets using rotations, horizontal and vertical flips, width and height shifts,
zoom and shear transforms. For that, we used Keras embedded tools within the
\texttt{ImageDataGenerator} module to augment images for the two-dimensional
networks. Since Keras's \texttt{ImageDataGenerator} is not able to process
three-dimensional input so far, we adapted the \texttt{ImageDataGenerator}
module. The adapted version we used in this study is named \texttt{ChunkDataGenerator},
and is available in the Supplementary Material.

To reduce the possibility of overfitting, we implemented dropout regularization
\cite{SRIVASTAVA:2014} in our pipeline. We followed the suggestions in the original
papers for U-net architectures: 2D U-net received a dropout rate of 50\% in the last
analysis layer and in the bottleneck, while 3D U-net \cite{CICEK:2016} did not
receive any dropout. The Tiramisu structures received a dropout rate of 20\%,
as suggested by Jégou et al \cite{JEGOU:2017}.

For a better comparison, we maintained the same training hyperparameters when possible.
Due to the large amount of training data and the similarities between training samples
(2D tiles or 3D cubes), our preliminary tests indicated that we would have a higher
accuracy for all networks in the first training epochs. Therefore, we decided to train
all architectures during five epochs. The 2D architectures were trained with batches
of four images, while the batches for 3D architectures had two cubes each. For all
architectures, we used a learning rate of $1E-4$, and binary cross entropy
\cite{ZHANG:2018} as the loss function. We followed the original papers regarding
to optimization algorithms: we used the Adam optimizer \cite{KINGMA:2017}
in the U-net architectures, while the Tiramisu ones were trained using the RMSProp
optimizer \cite{DAUPHIN:2015}. We implemented batch normalization
\cite{IOFFE:2015} in all architectures, including the 2D U-net. Ronneberger et
al. do not suggest it in their preliminary study, although it is known that
architectures using batch normalization tend to converge faster.

\subsection{Evaluation}
\label{subsec:eval}

We used Dice \cite{DICE:1945} and Matthews \cite{MATTHEWS:1975} correlation
coefficients (Equations \ref{eq:dice}, \ref{eq:matthews}]) to evaluate our
results, assuming that the fiber detections from \cite{LARSON:2019_dataset}
contain a reasonable gold standard.

\begin{equation}
    Dice = \frac{2 \times TP}{2 \times TP + FP + FN}
    \label{eq:dice}
\end{equation}

\begin{equation}
    Matthews = \frac{TP \times TN - FP \times FN}{\sqrt{(TP+FN)(TP+FP)(TN+FN)(TN+FP)}}
    \label{eq:matthews}
\end{equation}

Dice and Matthews coefficients receive true positive (TP), false positive (FP),
true negative (TN), and false negative (FN) pixels, which are determined as:

\begin{itemize}
    \item\textbf{TP:} pixels correctly labeled as being part of a fiber.
    \item\textbf{FP:} pixels incorrectly labeled as being part of a fiber.
    \item\textbf{TN:} pixels correctly labeled as background.
    \item\textbf{FN:} pixels incorrectly labeled as background.
\end{itemize}

TP, FP, TN, and FN are obtained when the prediction data is compared with a
certain gold standard, which in this study is Larson's semi-supervised segmentation
data \cite{LARSON:2019_dataset}.

\subsection{Visualization}

Imaging CMC specimens at high-resolution as Larson et al samples
\cite{LARSON:2019_dataset} leads to large datasets — each stack we used in this
paper has around 14 GB after the reconstruction, for example\footnote{The
exceptions are the registered versions of cured samples 232p3, 235p4 and 244p1,
with 11 GB each, and the sample 232p3 wet with around 6 GB.}.

Frequently, the specialist needs software to visualize the result of their data
collection, but most of them fail to produce meaningful graphs without
considering advanced image analysis and/or computational platforms with generous
amounts of memory. One may use Jupyter Notebooks \cite{KLUYVER:2016}, which enable
domain scientists to quickly probe specimens imaged with X-ray microCT during their
beamtime. For this reason, the figures in this paper are all generated on standard
laptops with no more than 16 GB of RAM, which is the typical computation system at
hand.

We used matplotlib \cite{HUNTER:2007} and ITK \cite{YOO:2002} (Fig
\ref{fig:3d_render}) to generate our figures. Despite our use of methods that
consider either global or local information, we designed protocols that allow
any user to visualize essential content from their experiments recorded as 3D
image stacks.

\section{DATA AVAILABILITY}

The supplementary data generated in this study is available at
\url{https://datadryad.org/stash/dataset/doi:10.6078/D1069R}, under a CC0 (public
domain) license.

\section{CODE AVAILABILITY}

The software we produced throughout this study is available at
\url{https://github.com/alexdesiqueira/fcn_microct/}, under a BSD license.

\section{ACKNOWLEDGEMENTS}

AFS would like to thank Sebastian Berg, Ross Barnowski, Silvia Miramontes, Ralf
Gommers, and Matt Rocklin for the discussions on fully convolutional networks,
their structure and different frameworks. This research was funded in part by
the Gordon and Betty Moore Foundation through Grant GBMF3834 and by the Alfred
P. Sloan Foundation through Grant 2013-10-27 to the University of California,
Berkeley.

\bibliographystyle{plain}
\bibliography{references}
\end{document}